\input phyzzx
\voffset = -0.6truein
\parskip = 1truepc
\hoffset=-0.3in
\hsize=6.8in
\rightline {October 1995}
\rightline {RHBNC-TH-95-1}
\title {SPACETIME DUALITY AND $\textstyle SU(n,1)\over\textstyle
SU(n)\otimes U(1)$
COSETS OF ORBIFOLD COMPACTIFICATION}
\author{W. A. Sabra*}
\footnote *{ e-mail: uhap012@vax.rhbnc.ac.uk.}
\address {Department of Physics,\break Royal Holloway and
Bedford New College,\break University of London,\break Egham,
Surrey, U.K.}
\abstract {The duality symmetry group of the cosets ${\textstyle SU(n,1)\over
\textstyle SU(n)\otimes U(1)}$,
which describe the moduli space of a two-dimensional subspace of an orbifold
model with $(n-1)$ complex Wilson lines moduli, is
discussed. The full duality group and its explicit action on the moduli fields
are derived.}
\endpage
\REF\one{L. Dixon, J. A. Harvey, C. Vafa and E. Witten, {\it Nucl. Phys.}
{\bf B261} (1985) 678; {\bf B274} (1986) 285.}
\REF\two{ A. Font, L. E. Ib\'a\~nez,
F. Quevedo and A. Sierra, {\it Nucl. Phys.} {\bf B331} (1991) 421.}
\REF\k{ J. Erler and  A. Klemm, {\it Comm. Math. Phys.} {\bf 153} (1993) 579.}
\REF\su{E. Cremmer, S.
Ferrara, L. Giraradello and A. Van Proeyen, {\it Nucl. Phys.}
{\bf B212} (1983) 413.}
\REF\verlinde{ R. Dijkgraaf, E. Verlinde and H. Verlinde,
{\it On Moduli Spaces of
Conformal Field
Theories with $c \geq 1$}, Proceedings Copenhagen Conference,
Perspectives
in String Theory,
edited by P. Di Vecchia and J. L. Petersen,
World Scientific, Singapore, 1988.}
\REF\yu{ L. Dixon, D. Friedan, E. Martinec and S. H. Shenker,
{\it Nucl. Phys.} {\bf B282} (1987) 13;
S. Hamidi and C. Vafa,  {\it Nucl. Phys.} {\bf B279} (1987) 465.}
\REF\dixon{L. Dixon, V. Kaplunovsky and J. Louis, {\it Nucl. Phys.}
{\bf B329} (1990) 27.}
\REF\luest{G. L. Cardoso, D. L\"{u}st and T. Mohaupt,  {\it  Nucl. Phys.}
{\bf B432} (1994) 68.}
\REF\lust{G. L. Cardoso, D. L\"{u}st and T. Mohaupt,  {\it  Nucl. Phys.}
{\bf B450} (1995) 115.}
\REF\nar{ K. S. Narain, {\it Phys. Lett.} {\bf  B169}  (1986) 41;
 K. S. Narain, M. H. Sarmadi and E. Witten, {\it Nucl. Phys.}
{\bf B279} (1987) 369.}
\REF\wilson{L. E. Ib\'{a}\~{n}ez, H. P. Nilles and F. Quevedo,
 {\it Phys. Lett.} {\bf B192} (1987) 332; L. E. Ib\'{a}\~{n}ez, J. Mas, H. P.
Nilles and F. Quevedo, {\it Nucl. Phys.} {\bf B301} (1988) 157;
T. Mohaupt, {\it  Int. J. Mod. Phys.} {\bf A9}  (1994)  4637.}
\REF\kahler{M. Cveti\v c, J. Louis and B. Ovrut, {\it Phys. Lett.} {\bf B206},
(1988) 229; M. Cveti\v c, B. Ovrut and W. A. Sabra, {\it Phys. Lett.} {\bf
B351} (1995) 173;
P. Mayr and S. Stieberger,  hep-th 9412196}
\REF\give{ A. Giveon, M. Porrati and E. Rabinovici,
{\it Phys. Rept}. {\bf 244} (1994) 77.}
\REF\Spa{M. Spalinski, {\it Phys. Lett.} {\bf B275} (1992) 47.}
\REF\waf{D. Bailin, A. Love, W. A. Sabra and S. Thomas,
{\it Phys. Lett.} {\bf B320} (1994) 21.}

\REF\modular{S. Ferrara, D. L\"ust, and S. Theisen,  {\it Phys. Lett.} {\bf
B233} (1989) 147;
S.  Ferrara, D. L\"ust,  A. Shapere and S. Theisen, {\it Phys. Lett.} {\bf
B225}(1989) 363 ; M. Cveti\v c,
A.  Font, L. E. Ib\'{a}\~{n}ez,  D. L\"ust and  F. Quevedo,
 {\it Nucl. Phys.} {\bf B361} (1991) 194;
S. Ferrara, C. Kounnas,  D. Lust and F. Zwirner,
{\it Nucl. Phys.} {\bf B365} (1991) 431.}
\REF\special{E. Cremmer and
A. Van Proeyen, {\it Class. and Quantum Grav.}  {\bf 2} (1985) 445;
E. Cremmer, C. Kounnas, A. Van Proeyen, J. P. Derendinger,
S. Ferrara, B. de Wit and L. Girardello, {\it Nucl. Phys.}
{\bf B250} (1985) 385;
B. de Wit and A. Van Proeyen, {\it Nucl. Phys.}
{\bf B245} (1984) 89;
L. Castellani, R. D'Auria and S. Ferrara, {\it Phys. Lett.}
{\bf B241} (1990) 57;
R. D'Auria, S. Ferrara and P. Fr\`e, {\it Nucl
Phys} {\bf B359} (1991) {705}.}
\REF\eleven{L. E. Ib\'{a}\~{n}ez, W. Lerche, D. L\"{u}st and S. Theisen,
{\it Nucl. Phys.} {\bf B352} (1991) 435; L. E. Ib\'{a}\~{n}ez and  D. L\"{u}st,
{\it Phys. Lett.} {\bf B302} (1993) 38.}
The derivation of a field theoretic low-energy effective action
from string theory is a first step in  attempting to  relate
string theory to the supersymmetric standard model or grand
unified theories. In particular, such a field theory should have
implications to fundamental questions in particle physics and cosmology, such
as the study of gauge coupling unification scale,
the masses of quarks and leptons and stringy
inspired inflationary scenarios.

Orbifold  compactified heterotic string theories are of great
phenomenological relevance. They constitute a large class of string vacua
where the interactions can be computed explicitly using the underlying
world-sheet conformal field theory [\yu].
The low-energy action is an $N=1$ supergravity coupled to Yang-Mills and
matter fields and their supersymmetric partners. With only terms with up to
two derivatives in the bosonic
fields, the theory is described in terms of three functions-
the K\"ahler potential $K$ encoding the kinetic terms for the
massless fields,  the superpotential $W$ containing the Yukawa
couplings and the gauge $f-$function whose real part, at the tree
level, determines the gauge couplings [\su]. The functions $K$ and $W$
appear in the Lagrangian of the theory  through  the combination
$${\cal G}=K+log\vert W\vert^2.\eqn\co$$
The orbifold models posses a set of continuous parameters,  the toroidal
moduli, parametrizing the size and shape of
the orbifold. The vacuum expectation values of the moduli
fields represent marginal deformations of the underlying conformal
field theory of the orbifold [\verlinde]. The toroidal moduli fields belong to
the untwisted sector of
the orbifold and enter the space-time $N=1$ supersymmetric
four-dimensional Lagrangian as chiral fields with flat
potentials to all orders in perturbation theory.
In addition to toroidal moduli, the untwisted sector of the
heterotic string theory compactified on orbifolds may also
contain Wilson lines moduli [\wilson].
These additional moduli exist in orbifold models where the twist
defining the orbifold is realized on the  $E_8\times E_8$ root
lattice by a rotation [\wilson].
Wilson line moduli are phenomenologicaly interesting because they
lower the rank of the gauge group and thus leading to more realistic models.
The moduli of the compactification on a $d$-dimensional torus
${\bf T}^d={\textstyle {\bf R}^d\over\textstyle\Lambda}$, where $\Lambda$ is a
$d$-dimensional lattice,
are encoded in the metric $G_{ij}$ which is the lattice metric of
$\Lambda$, an antisymmetric tensor $B_{ij}$ and
Wilson lines $A_{\ i}^I$, where $I$ is an $E_8\times E_8$ gauge lattice index
and
$i$ is an internal lattice index.
The moduli space of toroidal compactification [\nar] is given (locally)
by the coset space ${\textstyle SO(d+16,d)\over \textstyle SO(d+16,d)\otimes
SO(d)}.$
Toroidal compactifications lead to low-energy models with $N=4$
supersymmetry and gauge groups of rank $d+16$.
Six-dimensional orbifolds [\one,\two] are obtained by identifying the
points of the six-torus ${\bf T}^6$ under
a cyclic group $Z_N=\{ \theta^j, j=0,\cdots , N-1\}.$ In order to obtain
consistent space-time supersymmetric theories, the
twist should belong to $SU(3)$ but not $SU(2)$ [\one, \k]. Furthermore, to
reduce the
rank of the gauge group,  continuous Wilson line moduli must be introduced,
this can be achieved by allowing the orbifold twist to act on the
gauge sector of the theory as an automorphism of the $E_8\times E_8$
root lattice. It can be demonstrated [\luest, \kahler] that the moduli
spaces of orbifolds depend entirely on the eigenvalues of the twist and
their multiplicities.

The moduli space of the orbifold
are parametrized by the $T$ moduli corresponding to the K\"ahler
deformations  and the $U$ moduli which correspond to the
deformations of the complex structure. For each $U$ modulus, the corresponding
moduli space is  described by the coset
$\Big[{\textstyle SU(1,1)\over  \textstyle U(1)}\Big]$, and apart from
the ${\bf Z}_3$ orbifold
\foot{The moduli space of the $Z_3$ orbifold is
${\textstyle{SU(3,3)}\over \textstyle{SU(3)\otimes SU(3)\otimes U(1)}}$,
which is also special K\"ahler.},
the $T$ moduli spaces for all  symmetric orbifolds yielding $N=1$ space-time
supersymmetry are given by the special K\"ahler manifolds [\special]
$${\bf SK}(n+1)={SU(1,1)\over U(1)}\otimes {SO(n,2)\over SO(n)
\otimes SO(2)},\qquad n=2,4.$$
This structure can be derived using the Ward-identities of the
underlying world-sheet $(2,2)$ superconformal algebra [\dixon].

In the presence of Wilson lines moduli, the twist has more eigenvalues due to
the enlarged action of the twist on the $E_8\times E_8$ lattice and
the moduli spaces are given by  [\luest]
$$\bigotimes_{i=1}^{n} {SU(m_{i},n_{i})\over SU(m_{i})\otimes
SU(n_{i})\otimes U(1)}
\otimes {SO(p,q)\over SO(p)\otimes SO(q)},\eqn\kafka$$
where $m_i$($n_i$) and  $p$ ($q$) are the multiplicities of the complex
and $-1$ eigenvalues of the twist
on the left (right) moving sector, respectively.

A peculiar but phenomenologicaly interesting feature of  string
compactifications is that the physical
parameters of the low-energy effective theory are moduli dependent. Also, the
theory has  the target space duality
symmetry, the $T$ duality, which holds to all orders
in perturbation theory (see [\give] for a review). $T$-duality symmetries
consist of
discrete automorphisms of the moduli space which leave
the underlying conformal field theory invariant.
It is widely believed that non-perturbative effects in string
theory provide the mechanism for a range of unsolved problems,
namely, supersymmetry breaking,
lifting the vacuum degeneracy
in perturbative string theory and generating a non-trivial potential for
the dilaton field.
Also, non-perturbative potentials should have bearing on the questions
of cosmological inflation and the cosmological constant.
Presently, string theory is perturbative in its formulation and
physical principles by which non-perturbative physics
can be derived are not available. In this sense, duality symmetry play a
major role, if assumed to hold non-perturbatively, it puts strong
constraints on the form of any possible non-perturbative superpotential
in the four dimensional low-energy effective action [\modular].

It is our purpose here to derive the full duality symmetry and its action
on the moduli parameterizing  the cosets ${\textstyle SU(n,1)\over \textstyle
SU(n)\otimes U(1)}.$ These moduli spaces  appear in orbifold models where
Wilson lines moduli are present. So far only the action of a subgroup of the
duality symmetry on the moduli representing these cosets has been discussed
[\luest]. As a warm up exercise, we start by reviewing the duality symmetries
for the coset
${\textstyle SO(2, 2)\over \textstyle SO(2)\otimes SO(2)},$ describing the
moduli space of a two-dimensional
toroidal or ${\bf Z}_2$ orbifold without Wilson lines.
Then, duality symmetry for
two-dimensional ${\bf Z}_N$ orbifold or a two dimensional subspace of a
factorizable six-dimensional orbifold with a $Z_N$ twist ($N\not=2$),
 whose moduli space is given by
${\textstyle SU(1,1)\over \textstyle SU(1)},$ is  determined
in terms of both $SL(2)$ and $SU(1,1)$ groups.
In terms of $SL(2)$,
the duality group is given by all those elements with
integer values, $i. e.$, $SL(2,{\bf Z})$. However,
in terms of $SU(1,1)$, it will be demonstrated that the duality
group is not $SU(1,1, {\bf Z})$ but a subgroup whose elements  depend on the
particular twist defining the orbifold. These calculations are then
extended to the cosets ${\textstyle SU(n ,1)\over \textstyle SU(n)\otimes
U(1)},$
representing  the moduli spaces of two-dimensional subspaces
of orbifold compactification where continuous Wilson lines are present and
where the twist has a complex eigenvalue.
The full duality group and its action on the moduli fields are derived.
Finally, we comment on the duality transformations of the basic physical
parameters defining the low-energy effective action.

Duality symmetry is a discrete symmetry acting on the the moduli
space of the underlying conformal field theory.
The action on the moduli is such that the underlying conformal
field theory is invariant. The vertex operators of the
underlying conformal theory depend both on the moduli and a set of
quantum numbers, the winding, momenta and gauge quantum numbers, this
implies that duality symmetry has
a non-trivial action on the quantum
numbers in order to have a duality invariant spectrum.
We will discuss the duality symmetry of a two-dimensional torus and
its corresponding ${\bf Z}_N$ orbifolds.

The two-dimensional toroidal compactification is described by four real
parameters
represented  by
the independent components of the
antisymmetric tensor $B_{ij}$, and  the lattice metric $G_{ij}.$
The vertex operators of the underlying conformal field theory
have the following spins and scaling dimensions
(ignoring the oscillators contribution)
$$H= {1\over2} (P^t_L G^{-1}P_L + P^t_R G^{-1}P_R), \qquad
S= {1\over2} (P^t_L G^{-1}P_L - P^t_R G^{-1}P_R),\eqn\latt$$
where the left and right momenta are given by
$$P_L={p\over 2}+(G-B)w, \qquad P_R={p\over2}-(G+ B)w,\eqn\latt$$
here the index $t$ denotes the transpose, $w$ and $p,$ the windings and
momenta, respectively,
are two-dimensional integer valued vectors taking values on
the two dimensional lattice $\Lambda$ of the torus and its dual $\Lambda^*$ and
 $G$ and $B$ are $2\times 2$ matrices representing the background metric and
antisymmetric tensor.

To identify the symmetries of the spectrum, it is more convenient to
write $H$ and $S$ in a matrix form [\Spa]
$$H={1\over2}u^t\Xi u,\qquad S={1\over2}u^t\eta u,\eqn\andy$$
where
$$\Xi=\pmatrix{2(G-B)G^{-1}(G+B)&BG^{-1}\cr
-G^{-1}B&{1\over2}G^{-1}},\qquad
\eta =\pmatrix{{\bf 0}&{\bf 1}\cr {\bf 1}&{\bf 0}},
\qquad u=\pmatrix{n_1\cr n_2\cr m_1\cr m_2}.\eqn\de$$
In terms of the complex moduli
defined as,
$$T=T_1+iT_2=2({\sqrt{\det G}}-ib),\qquad
U=U_1+iU_2={1\over G_{11}}({\sqrt{\det G}}-iG_{12}),\eqn\com$$
$\Xi$ is expressed by
$$\Xi={1\over {T_1U_1}}\pmatrix{\vert T\vert^2&-\vert T\vert^2
U_2&-T_2U_2&-T_2\cr
-\vert T\vert^2 U_2&\vert T\vert^2\vert U\vert^2&T_2\vert U\vert^2&
T_2U_2\cr -T_2U_2&T_2\vert U\vert^2&\vert U\vert^2&U_2
\cr -T_2&T_2U_2&U_2&1},\eqn\mary$$
and
$$H=\vert {\bf P}_R\vert^2+S=2{\vert m_2-im_1U+in_1T-n_2UT\vert^2
\over(T+\bar T)(U+\bar U)}+S.\eqn\natasha$$
The discrete target space duality symmetries are defined by
all integer-valued linear transformations of the quantum numbers which
leave the spectrum invariant, this means that  the duality transformations
must leave both $S$ and $\vert {\bf P}_R\vert^2$ invariant.
Denote such a transformation by $\Omega$ with an  action on the
quantum numbers defined by
$$u\rightarrow \Omega^{-1}u.\eqn\tra$$
It is then clear that for $S$ to be invariant, $\Omega$ must satisfy
$$\Omega^t\eta\Omega=\eta.\eqn\al$$
Also the requirement of the invariance of $H$ under the action of $\Omega,$
defines the duality transformation of the moduli fields. This is given by
$$\Xi\rightarrow \Omega^t\Xi\Omega.\eqn\zozo$$
In terms of the $T$ and $U$ parametrization of the moduli space,
the duality symmetry can be represented  by
$SL(2,{\bf Z})_T\otimes SL(2,{\bf Z})_U\otimes {\bf Z}_2^{(1)}\otimes
{\bf Z}_2^{(2)}$
which act on the moduli as
$$\eqalign{SL(2,{\bf Z})_T: \quad T&\rightarrow {aT-ib\over icT+d},
\quad U\rightarrow U \qquad ad-bc=1,\cr
SL(2,{\bf Z})_U: \quad U&\rightarrow {a'U-ib'\over ic'U+d'},\quad
T\rightarrow T \qquad a'd'-b'c'=1,\cr
{\bf Z}_2^{(1)}:\quad  T&\leftrightarrow U,\cr
{\bf Z}_2^{(2)}:\quad T&\leftrightarrow \bar T, \quad U\leftrightarrow
\bar U.}\eqn\ascot$$
The $SL(2, {\bf Z})_T$ and $SL(2, {\bf Z})_U$ action on the quantum
numbers is represented by [\waf]
$$\Omega_T=\pmatrix{a&0&0&c\cr 0&a&-c&0\cr 0&-b&d&0\cr b&0&0&d},\qquad
\Omega_U=\pmatrix{d'&b'&0&0\cr c'&a'&0&0\cr 0&0&a'&-c'\cr 0&0&-b'&d'}.
\eqn\aaa$$
The generalization of the above results to the orbifold is
straightforward. Define the action of the twist on the quantum numbers by
$$u \longrightarrow u^\prime = {\Theta}u,\qquad {\Theta}= \pmatrix{
	Q & 0 \cr
	0& {(Q^t)}^{(-1)}} \qquad {\Theta}^N=1,\eqn\qu$$
here $Q$ is an integer-valued matrix and $N$ is the order of the twist.
For the twist to act as a  lattice automorphism, the background fields
must satisfy the conditions
$${\Theta}^t\Xi {\Theta} = \Xi,\quad \Rightarrow\qquad
Q^tGQ=G,  \qquad Q^tBQ=B.\eqn\comp$$
Clearly, the conditions \comp\ imply that the orbifold has less
moduli than its corresponding torus.
The duality symmetry of the orbifold are those of the
corresponding torus satisfying the additional condition
${\Theta}\Omega=\Omega{\Theta}^k$
with $1\le k\le N$ [\luest, \Spa]. It is obvious that a  two-dimensional ${\bf
Z}_2$ has
the same duality symmetry  as that of the corresponding torus.
However, for a ${\bf Z}_N$ twist, with $N\not=2$, the $U$ modulus
is frozen to a constant complex value,
and the duality symmetry is for two-dimensional ${\bf Z}_N$ orbifolds
is  $SL(2,{\bf Z})_T$. This can be easily from eq. \natasha\ after setting $U$
to a constant value $U_0.$

In order to identify the duality symmetry group of the coset
$\textstyle SU(1,1)\over\textstyle U(1)$ in terms of the group
$SU(1,1)$, a  complex basis $u_c$ for the quantum numbers
is required in which  the spin is given by an $SU(1,1)$ quadratic form.
A way to get this new basis of the quantum numbers is to use
a different parametrization of the moduli space. If one writes
$U_0=u_1+iu_2$, and perform the change of variables
$t={\textstyle{1-T'}\over\textstyle{1+T'}},$ with
$T'={\textstyle T\over\textstyle 2u_1}$, then from \natasha\ we obtain
$$\vert {\bf P}_R\vert^2=2{\vert m_c-n_ct\vert^2\over (1-t\bar
t)}=u_c^\dagger\Xi_c u_c,\qquad S=u^\dagger _c L u_c\eqn\st$$
where\foot{ expressions for $m_c$ and $n_c$ have been obtained  in [\lust] for
the two dimensional ${\bf Z}_3$ orbifold case using a different approach. The
expressions in (20) are valid for all orbifolds.}
$$\eqalign{&L=\pmatrix {1&0\cr 0&-1},\quad u_c=\pmatrix{n_c\cr m_c},\quad
\Xi_c={2\over 1-t\bar t}\pmatrix{t\bar t&-\bar t\cr -t&1},\cr
& m_c={1\over{2\sqrt2}u_1}(m_2-im_1U_0+2in_1u_1-
2n_2u_1U_0),\cr
& n_c={1\over{2\sqrt2}u_1}(-m_2+im_1U_0+2in_1u_1-2n_2u_1U_0).}
\eqn\fraser$$

An element of $SU(1,1)$ represented by $\Omega_c,$ is a $2\times 2$ complex
matrix with unit determinant satisfying
$$\Omega_c^{\dagger}L\Omega_c=L,\qquad
\Omega_c=\pmatrix{z_1&z_2\cr z_3&z_4}.\eqn\dd$$
The condition in \dd\ implies
$$\bar z_1=z_4,\qquad\qquad\bar z_2=z_3.\eqn\ste$$
If we define the action of $\Omega_c$ on the quantum numbers by
$$u_c\rightarrow\Omega^{-1}_cu_c,\eqn\stut$$
then clearly this action leaves $S$ invariant. For $\vert P_R\vert^2$ to remain
invariant under
the transformation \stut, the moduli field should transform as follows
$$\Xi_c\rightarrow
\Omega_c^\dagger\Xi_c\Omega_c.\eqn\shine$$
{}From \shine\  one can extract the duality transformation for the moduli $t$.
This is given by
$$t\rightarrow {z_1t-z_3\over z_4-z_2t}={z_1t-\bar z_2\over \bar z_1-z_2t};
\quad \vert z_1\vert^2-\vert z_2\vert^2=1.\eqn\raj$$
The elements of $SU(1,1)$ transformations can be expressed in
terms of those of $SL(2,{\bf Z})$ by
$$\eqalign{z_1=\bar z_4=&{1\over2}(a+d)+{i\over2}({b\over2u_1}-2u_1c),\cr
z_2=\bar z_3=&{1\over2}(a-d)+{i\over2}({b\over2u_1}+2u_1c).}\eqn\para$$
We stress here that the duality group is not $SU(1,1,{\bf Z})$ but a subgroup
of $SU(1, 1)$ whose elements depend on the particular orbifold ($u_1$ is
different for different orbifolds).

We now turn to discuss the full
duality symmetry of the moduli space of a two-dimensional subspace of a
factorizable six-dimensional orbifold, where Wilson lines are present and where
the twist is given by $Z_N$ ($N\not=2$). A subgroup of this duality symmetry
and its action on the moduli has been discussed in [\luest]. For simplicity we
will consider Wilson line modulus with only two gauge indices. Generalization
to more than two gauge indices is straightforward. Like $G$ and $B,$ the Wilson
line modulus also  satisfies consistency requirement. If we represent the
Wilson line components,  in an orthonormal basis of the
gauge lattice, by the matrix
$$A=\pmatrix{A_{\ 1}^1&A_{\ 2}^1\cr A_{\ 1}^2&A_{\ 2}^2},\eqn\maha$$
then consistency condition implies
$$MA=AQ,\eqn\trieste$$
where $M$ defines the action of the twist on the gauge quantum numbers in the
same way $\Theta$ defines the action of the twist on the winding and momenta.
Using the methods described in the previous section, we obtain
[\lust]
$$\eqalign{&\vert {\bf P}_R\vert^2={\vert in_1T-n_2U_0T-im_1U_0+m_2+
{i u_1}Q_1{\bf A}- Q_2u_1{\bf A}\vert^2\over
{1\over2}(U_0+\bar U_0)(T+\bar T)-(u_1^2 A\bar A)},\cr
& S={1\over2}(Q_1^2+Q_2^2)+m_1n_1+m_2n_2.}\eqn\rig$$
Here  $Q_1$ and $Q_2$ are the gauge quantum numbers
with respect to an orthonormal gauge lattice basis, and
the complex moduli $T$ and ${\bf A}$ are given in terms of the
real components of the Wilson line and the metric by
$$\eqalign{&{\bf A}=2u_1(A^1_{\ 1}-iA^2_{\ 1}),\cr
& T=2\Big(\sqrt{\det G}(1+{1\over4}{A_{\ 1}^aA_{a1}\over G_{11}})-
i(b+{1\over4}{A_{\ 1}^aA_{a1}G_{12}\over G_{11}}-
{1\over4}A_{\ 1}^aA_{a2})\Big)}\eqn\restless$$
and $U_0=u_1+iu_2$ is the fixed value of the $U$ modulus.
Due to the consistency conditions \comp\ and \trieste, the following conditions
must be satisfied
$$\eqalign{A^2_{\ 2}=&u_1 A^1_{\ 1}-u_2 A^2_{\ 1},\cr
A^1_{\ 2}=&-u_2 A^1_{\ 1}-u_1 A^2_{\ 1},\cr
G_{12}=&-u_2 G_{11},\cr
G_{22}=&\vert U_0\vert^2G_{11}.}\eqn\rel$$
The moduli $T$ and $\bf A$ parametrize the moduli space ${\textstyle
SU(2,1)\over \textstyle SU(2)\otimes U(1)}.$ Again in order to identify the
duality group in terms of
the group $SU(2,1)$, we perform the following change of variables,
$${T\over 2u_1}={1-t\over 1+t},\quad {\bf A}=2{{\cal A}\over1+t}.\eqn\cruise$$
In terms of the new parametrization \cruise\ of the moduli
space, $\vert {\bf P}_R\vert^2$ and $S$ in \rig\ can be written as
$$\vert {\bf P}_R\vert^2=2{\vert Q_c{\cal
A}-n_ct+m_c\vert^2\over (1-t\bar t-{\cal A}\bar {\cal A})}=v_c^\dagger\xi_c
v_c,\qquad
S=v^\dagger_c{\hbox {L}}v_c.\eqn\Zepa$$
where $$\eqalign{& v_c=\pmatrix{Q_c\cr n_c\cr  m_c},\qquad {\hbox
{L}}=\pmatrix{ 1& 0&0\cr
 0& 1& 0\cr  0& 0&-1},\qquad Q_c={1\over\sqrt2}(iQ_1-Q_2),\cr
& \xi_c={2\over 1-t\bar t-{\cal A}\bar{\cal A}}\pmatrix{{\cal A}\bar{\cal
A}&-t\bar {\cal A}&\bar {\cal
A}\cr -\bar t{\cal A}&t\bar t&-\bar t\cr {\cal A}&-t&1}}\eqn\holi$$
and $m_c$, $n_c$ are as given in \fraser.

An element of $SU(2,1),$ $\Gamma,$ is a $3\times 3$ matrix with unit
determinant satisfying
$$\Gamma=\pmatrix{ z_1&z_2&z_3\cr
z_4&z_5&z_6\cr z_7&z_8&z_9},\qquad
\Gamma^\dagger {\hbox {L}}\Gamma={\hbox {L}}.\eqn\suto$$
The action of $\Gamma$  on the complex quantum numbers and moduli can be
represented by
$$v_c\rightarrow \Gamma^{-1} v_c,\qquad \xi_c\rightarrow
\Gamma^\dagger\xi_c\Gamma.\eqn\nato$$
This gives the following transformations for $t$ and $\cal A$,
$$\eqalign{t&\rightarrow {\tilde t}={z_5t-z_2{\cal A}-z_8\over z_3{\cal
A}+z_9-z_6t},\cr
{\cal A}&\rightarrow {\tilde {\cal
A}}={z_1{\cal A}-z_4t+z_7\over z_3{\cal A}+z_9-z_6t}.}
\eqn\nic$$
Moreover, in addition to \suto\ there are further constraints
on the elements of $\Gamma$,  these conditions arise from the fact the physical
quantum numbers should transform as integers under the duality transformations.

As an illustration, consider the ${\bf Z}_3$ orbifold where  the internal
lattice and the gauge lattice are both given by the root lattice of $SU(3)$ and
the twist is defined by the
Coxeter action. In this case, the action of the twist on the winding
numbers and quantum gauge numbers is given by
$$Q=\pmatrix{0&-1\cr 1&-1},\qquad M={1\over2}\pmatrix{-1&-\sqrt3\cr
\sqrt3&-1},\eqn\kjl$$
where  an orthonormal basis for the gauge lattice has been chosen. The
orthonormal vectors $E^a$, $a=1, 2$, are given
in terms of the $SU(3)$ root lattice
vectors $e^I$ by
$$\eqalign{e^1=&{\sqrt2} E^1,\cr e^2=&-{1\over\sqrt2}E^1+
{\sqrt{3\over2}}E^2.}\eqn\houtaf$$
Substituting \kjl\ in \comp\ and \trieste\ and solve for the background fields
$G$  and $A$ we get
$$\eqalign{& G_{12}=-{1\over2}G_{11},\qquad  G_{22}=G_{11},\cr
& A_{\ 2}^1=-{1\over2} A_{\ 1}^1-{\sqrt3\over2}A_{\ 1}^2,\qquad
 A_{\ 2}^2={\sqrt3\over2}A_{\ 1}^1-{1\over2}A_{\ 1}^2.}\eqn\asder$$

The generalization of the above calculations to the
${\textstyle SU(n, 1)\over\textstyle SU(n)\otimes U(1)}$ cases where  one has
$n-1$ complex Wilson lines is straightforward. In these cases, the Wilson line
and $Q_c$  in \Zepa\ will have indices ranging from 1 to $n-1$ and the duality
symmetry in terms of $SU(n, 1)$ can be derived along the same lines discussed
for the case of $SU(2, 1)$.

We now discuss the action of duality symmetry on the tree level low-energy
Lagrangian. The K\"ahler potential for the coset $\textstyle SU(2,1)\over
\textstyle SU(2)\otimes U(1)$ coset
is given by [\luest, \kahler]
$$K=-log(1-t\bar t-{\cal A}{\bar{\cal A}}).$$
Using the conditions which arise from \suto\
$$\eqalign
{&\bar z_1=z_5 z_9-z_6z_8,\qquad
\bar z_4=z_3z_8-z_2z_9,\qquad
\bar z_7=z_3z_5-z_2z_6,\cr &\bar z_2= z_6z_7-z_4z_9,\qquad
\bar z_5=z_1z_9-z_3z_7,\qquad\bar z_8=z_1z_6-z_3z_4,\cr
&\bar z_3= z_5z_7- z_4z_8,\qquad
\bar z_6= z_1z_8- z_2 z_7,\qquad
\bar z_9= z_1z_5- z_2 z_4,}\eqn\suton$$
it can be easily seen  that the duality transformations \nic\
induce a  K\"ahler transformation on the K\"ahler potential,
$$K\rightarrow K+log ( z_3{\cal A}+z_9-z_6t)+
log ( \bar z_3\bar{\cal A}+\bar z_9-\bar z_6\bar t).\eqn\nora$$
As was mentioned in the introduction, ignoring gauge terms, the
low-energy effective Lagrangian is described in terms of the function
$\cal G$ defined in eqn \co. Therefore for the low-energy theory to be
invariant under the duality transformation \nic\ associated with the coset
${\textstyle SU(2,1)\over \textstyle SU(2)\otimes U(1)},$  the superpotential
must transform as (up to a
${\cal A},$
$t$-independent phase)
$$W\rightarrow W(z_3{\cal A}+z_9-z_6t)^{-1}.\eqn\nnn$$
Eq. \nnn\ provides a non-trivial constraint on the form of any possible
non-perturbative superpotential.  The
duality transformations for the various fields in the low-energy effective
action can be determined from the form of their associated vertex operators
[\eleven].  It would be interesting to investigate the duality transformations
of the twist fields and their interactions  and verify that \nnn\ holds for the
Yukawa couplings in the twisted sectors.

To summarize, we have analyzed the duality structure of the coset
spaces ${\textstyle SU(n,1)\over \textstyle SU(n)\otimes U(1)}.$ These spaces
describe the moduli spaces of a two-dimensional torus in a factorizable
orbifold where the twist has a complex eigenvalue and are parametrized by  a
complex
K\"ahler moduli $\bf T$  and $n-1$ complex Wilson moduli $\bf A.$ Also, the
duality transformations of the basic physical parameters in the low-energy
effective action are discussed.
\vskip 1cm
\centerline{\bf ACKNOWLEDGEMENT}I  thank S. Thomas and Jose Figueroa-O'Farrill
for useful
comments. This work is supported by P.P.A.R.C.
\vfill\eject
\refout
\end